\documentclass[aps,pra,twocolumn,groupedaddress,superscriptaddress]{revtex4}
\usepackage{amsmath}\usepackage{xcolor}\usepackage{epstopdf}
\usepackage{graphicx}
\usepackage{color,ulem} 
\usepackage{bm} % bold math symbols

\draft % marks overfull lines with a black rule on the right

\usepackage{color}
\usepackage{amsmath}
\usepackage{ulem}
\usepackage{gensymb}

\begin{document}

\title{{Tailoring optical pulling forces with composite microspheres}} %Title of paper

\author{R. Ali}
\email[]{r.ali@if.ufrj.br}
\affiliation{%
Instituto de F\'isica, Universidade Federal do Rio de Janeiro, Caixa Postal 68528, Rio de Janeiro, RJ, 21941-972, Brasil
}
\author{F. A. Pinheiro}
\affiliation{%
Instituto de F\'isica, Universidade Federal do Rio de Janeiro, Caixa Postal 68528, Rio de Janeiro, RJ, 21941-972, Brasil
}
%\author{ F. S. S. Rosa}
%\affiliation{%
%Instituto de F\'isica, Universidade Federal do Rio de Janeiro, Caixa Postal 68528, Rio de Janeiro, RJ, 21941-972, Brasil
%}
\author{ R. S. Dutra}
\affiliation{LISComp-IFRJ, Instituto Federal de Educa\c{c}\~ao, Ci\^encia e Tecnologia, Rua Sebasti\~ao de Lacerda, Paracambi, RJ, 26600-000, Brasil}
\author{P. A. Maia Neto}
\affiliation{%
Instituto de F\'isica, Universidade Federal do Rio de Janeiro, Caixa Postal 68528, Rio de Janeiro, RJ, 21941-972, Brasil
}

\begin{abstract}

Optical pulling forces or tractor beams can pull particles against light propagation by redirecting the incident photons forward. This is typically achieved using 
 Bessel beams with very small half-cone angles, which considerably limits its applicability. 
One can circumvent such issue by using
 a superposition of plane waves. In order to investigate optical pulling forces 
 exerted by  a pair of non-colinear plane waves, 
  we develop a theoretical framework based on Mie theory, Debye potentials and Wigner rotation matrices.
   We apply this framework to calculate the optical pulling force on metallo-dielectric composite particles, which we put forward as an alternative material platform to optimize and tailor tractor beams. Indeed we demonstrate that by adding a few plasmonic inclusions to low-refractive index dielectric particles of arbitrary sizes, we are able to produce polarization dependent optical pulling forces that cannot occur in the corresponding homogeneous particles. Altogether our findings not only provide innovative theoretical methods to compute optical pulling forces, but also provide new strategies to tailor and optimize them, paving the way to increase their applicability.
\end{abstract}

%\setboolean{displaycopyright}{true}

%\begin{document}

\maketitle

%\thispagestyle{fancy}
%\ifthenelse{\boolean{shortarticle}}{\abscontent}{}

\section{Introduction}
 Light exerts radiation pressure on  matter due to momentum exchange.
 Focused light beams exert an optical scattering force (forward force due to radiation pressure),  as well as a  gradient force (possibly along the backward direction)
  on small particles owing to the inhomogeneity  of the electromagnetic field.
 As an example of application, in optical tweezers a tightly focused beam is used for optical manipulation \cite{Ashkin1970,ashkin1986,ashkin2006,Gieseler2020} and optical rotation \cite{Friese1998,grier2003,Lin2014,Diniz2019,Ali2020} with many applications \cite{fazal2011,Brzobohaty2015,Marago2018,Polimeno2018}.
 
 {On the other hand, a plane wave will always push a particle made of a passive medium along the forward direction.} 
However, under certain conditions 
{spatially structured beams}
 can accelerate small particles {along the direction opposite} to the light propagation direction
\cite{Lee2010,chan2011,Shvedov2014,chen2015,li2019}.
Negative optical force or optical pulling force (OPF)  occurs whenever illuminated particles are pulled towards the source due to the momentum conservation.
In contrast to the trapping force in optical tweezers, the  OPF can accelerate particles
over a long distance 
 without defining an equilibrium position.
 OPF has attracted considerable attention due to  its many applications  such as optical sorting \cite{Zhu2018,bowman}, self-assembling,  remote sampling \cite{Brzobohaty2013,Mazilu2012},  miniaturization of nanodevices~\cite{lipson2009}
 and enantioselective manipulation \cite{Canaguier2013,Kun2014,bradshaw2014,wang2014,hayat2015,Kamandi2017} {(see \cite{Ding2019} for a recent review).} 
 In particular, it was both theoretically and experimentally shown that  nondiffracting  Bessel beams can pull a sub-wavelength dielectric sphere \cite{chan2011,Brzobohaty2013,mitri2016}. Several approaches exist to demonstrate OPF on homogeneous dielectric spherical particles using bichromatic fields \cite{Krasnov2012}, multiple plane waves \cite{Liu2017,Mobini2018} and with  optical gain \cite{Bian2017,Alaee2018}. 

OPF can be achieved for weakly absorptive particles that maximize forward scattering while minimizing backward scattering~\cite{chan2011,Liu2017,dogariu2011,Ding2019}.
Typically, it is easier to pull high-refractive index particles, and even more so in the size range close to the laser wavelength $\lambda.$
Particles made of Si, Ge and GaAs can be pulled even when very small, with radii in the range $r\stackrel{>}{\scriptscriptstyle\sim}0.2\,\lambda$
 because of the significant forward scattering and negligible backscattering due to the strong coherent interference between electric and magnetic dipoles~\cite{Liu2017,Mobini2018}.
  In contrast,
particles made of  lower refractive index materials such as polystyrene or silica
 can only be pulled for relatively larger radii $r\stackrel{>}{\scriptscriptstyle\sim}0.4\,\lambda$~\cite{chan2011,Brzobohaty2013}.

The angular distribution of Mie scattering is clearly a key feature to achieve OPF. 
 In that respect, 
  progress in the fields of metamaterials and plasmonics now allows for novel strategies to tailor
   the interplay between electric and magnetic multipoles  to produce directional Mie scattering~\cite{chan2011,Liu2017,kall2018}.
In particular, when considering composite microspheres, it is possible to
 minimize backscattering by tuning the filling fraction of inclusions \cite{Ali2018,Ali2020}.

In this work we show that metallo-dielectric composite microspheres, with metallic inclusions embedded in a low-index dielectric host, provide an optimal material platform 
that allows for pulling forces in a size range well below the laser wavelength,  thus extending the 
 applicability domain of OPF towards smaller sizes and lower refractive indexes. 
 Specifically, we put forward a scheme using a superposition of plane waves, in contrast to the traditional approach that employs structured light beams. 
Our model describes the case of two collimated Gaussian
 light beams with beam waists much larger than the particle size thus leading to long-ranged optical forces. This is typically the case of paraxial beams as we consider 
 sphere radii smaller or of the order of $\lambda.$
 In order to consider our proposal, we develop 
 a novel analytical framework to compute optical forces based on Mie theory, Debye potentials and Wigner rotation matrices, and apply it to situations such as the one depicted in Fig. \ref{fig:a1}.

This paper is organized as follows. Sec. 2 is devoted to the derivation of our
theoretical formalism, which is based on Mie scattering theory combined with Wigner rotations. 
In Sec.~3, we present numerical results for the OPF on a metallo-dielectric composite. 
 Finally,  we summarize our findings and conclusions in Sec.~4. 

\section{Optical force exerted by a superposition of plane waves on a Mie sphere }

\begin{figure} 
\includegraphics[width = 3.2in]{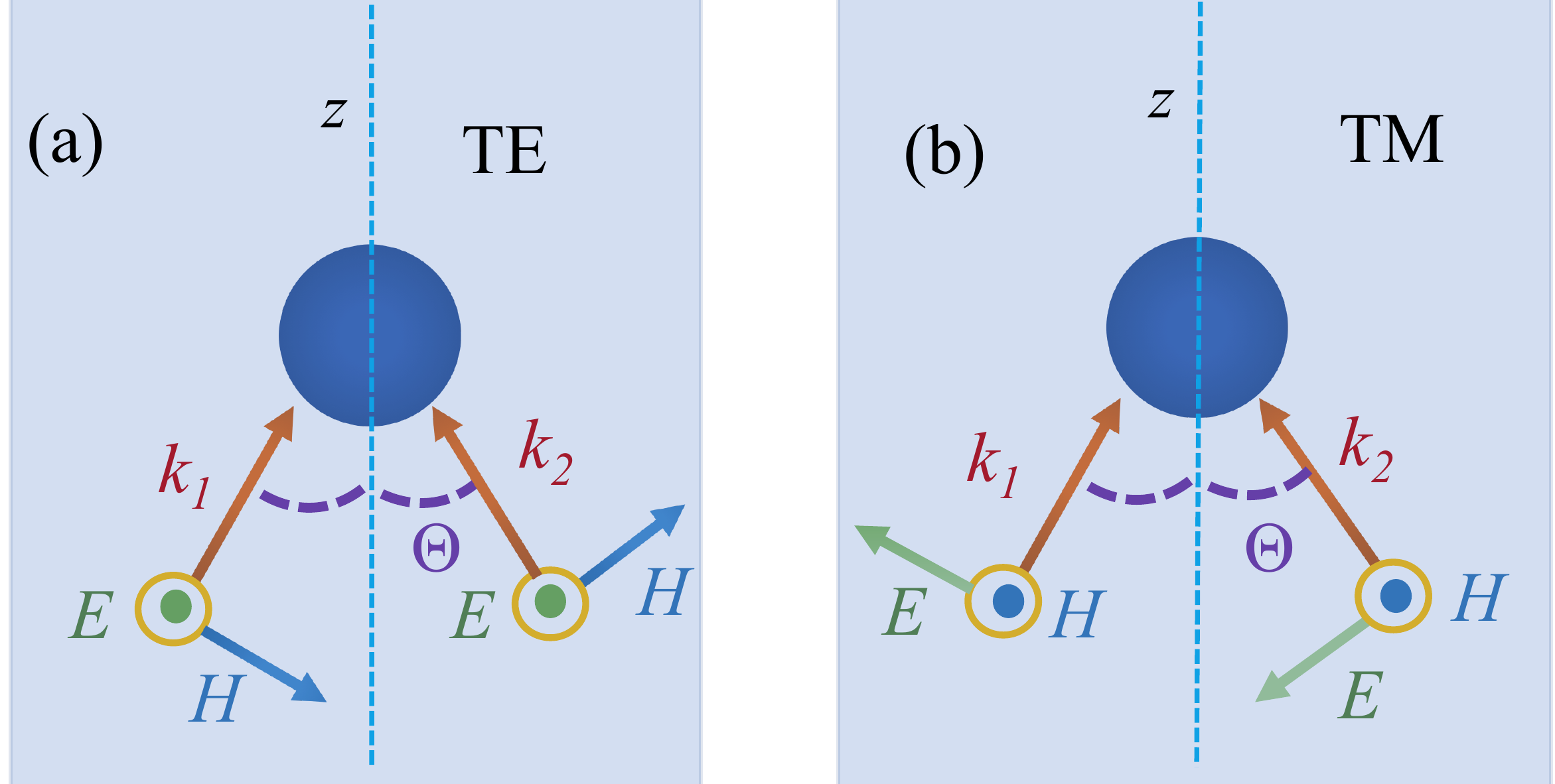}
\caption{ Schematic diagram of a particle illuminated by two non-colinear plane waves with (a)  transverse electric  (TE)
and (b) transverse magnetic (TM) polarization. The angle between the corresponding wavevectors is $2\Theta.$}  
\label{fig:a1}
\end{figure}

In the following we present our theoretical approach to 
 the optical force on a microsphere. 
 We consider a superposition of two linearly-polarized  non-colinear plane waves, as indicated in Fig.~\ref{fig:a1}. 
 The corresponding wavevectors define the plane shown in Fig.~\ref{fig:a1}, with respect to which we define 
 transverse electric (TE) and transverse magnetic (TM) polarizations. We assume that both plane waves are  
 linearly polarized along the same direction. 
 The  mixed case or  a higher number of plane waves can also be considered within our formalism. 
In order to allow for arbitrary values of $R/\lambda,$ 
 our approach is based on Mie scattering, 
 implemented in terms of 
Debye potentials, 
and on Wigner rotations to build the desired scattering problem from 
the standard case of axial incidence. 

We start by writing the electric field corresponding to the superposition of two incident plane waves  
propagating in the non-magnetic and non-absorbing host medium: 
\begin{equation}
\mathbf{E}_{\rm in}({\bf r},t)=  \sum^{2}_{j=1}\, E_0\, \hat{\bm{\varepsilon}}_j\,e^{i( \mathbf{k}_{j}\cdot \mathbf{r}-\omega t)},  \label{incidentfield}
\end{equation}
where the polarization unit vectors ($j=1, 2$) are 
\begin{equation}\label{unitvectors}
\hat{\bm{\varepsilon}}_j(\theta_j,\phi_j)= 
\cos\phi_j\;\hat{\bm{\theta}}_j - \sin\phi_j\; \hat{\bm{\phi}}_j\,.
\end{equation} 
 The wavevectors 
\[
 {\bf k}_j = n_h\,\frac{\omega}{c}\, \left(\sin\theta_j \cos\phi_{j},\sin\theta_{j} \sin\phi_{j}, \cos\phi_{j} \right).
\]
 are written in terms of
 the speed of light in vacuum $c$ and of 
  the host medium refractive index $n_h=\sqrt{\epsilon_h},$ where 
 $\epsilon_h$ is the relative permittivity. 
The spherical polar and azimuthal angles $\theta_j$ and $\phi_j$ 
define the propagation directions with respect to the $z$-axis.
When discussing specific examples, we will take
  $\phi_{1}=0$ and $\phi_{2}=\pi$  for TM   and  
    $\phi_{1}=\pi/2$ and $\phi_{2}=3\pi/2$  for TE polarization as depicted in Fig. \ref{fig:a1}.

 We introduce the electric (E) and magnetic (M) Debye potentials~\cite{Bowkamp1954,Bobbert1986} 
\begin{eqnarray}
\label{PiE}
\Pi^{E} ({\bf r})&=& \sum^{\infty}_{\ell=1}\frac{({\bf r}\cdot {\bf E})_{}}{{\ell}({\ell}+1)}\\
\label{PiM}
\Pi^{M}({\bf r})& = &\sum^{\infty}_{\ell=1}\frac{({\bf r}\cdot {\bf H})}{{\ell}({\ell}+1)} \, . 
\end{eqnarray}
The electric and magnetic fields can be written in terms of the Debye potentials as follows:
\begin{eqnarray}
\label{EfromPi}
{\bf E}&=  & \nabla \times \nabla \times \left({\bf r}\, \Pi^{E}\right) + i\omega \mu_0 \,  \nabla \times\left({\bf r}\, \Pi^{M}\right)\\
\label{HfromPi}
{\bf H}&=&  \nabla \times \nabla \times\left({\bf r}\, \Pi^{M}\right) - i\omega \epsilon_h\epsilon_0\,  \nabla \times\left({\bf r}\, \Pi^{E}\right) \, ,
\end{eqnarray} 
where  $\epsilon_0$ and $\mu_0$ are the vacuum permittivity and permeability, respectively.

In order to derive the Debye potentials for a given plane wave, 
we first consider 
a `primed' coordinate system such that the $z'$ and $x'$ axes coincide with the propagation and polarization directions, respectively. 
In such system, we find (omitting the time dependence from now on)
\begin{equation}\label{start}
{\bf \hat{r}}\cdot {\bf E}(r,\theta',\phi')= E_0\, e^{ik r \cos\theta'} \, \sin\theta' \cos\phi'.
\end{equation}

The Debye potentials of a plane wave in the primed coordinate system are then obtained from Eqs.~(\ref{PiE}), (\ref{PiM}) and (\ref{start})
by expanding $e^{ik r \cos\theta'}$ in terms of spherical waves:
\begin{eqnarray}\label{PiEPlane}
\Pi_{\rm pw}^{E}(r,\theta',\phi') &=  & \frac{E_{0}}{k}  \sum^{\infty}_{\ell=1}
 i^{\ell+1} \sqrt{\frac{\pi(2\ell+1)}{\ell(\ell +1) }}  j_{\ell}(kr)\\
 &&\times \Bigl[Y_{\ell,+1}(\theta' , \phi')-Y_{\ell,-1 }(\theta' , \phi')\Bigr]\nonumber
\end{eqnarray}

\begin{eqnarray}\label{PiMPlane}
\Pi_{\rm pw}^{M}(r,\theta',\phi')&=&  \frac{H_{0}}{k}   \sum^{\infty}_{\ell=1} i^{\ell} \sqrt{\frac{\pi(2\ell+1)}{\ell(\ell +1) }}  j_{\ell}(kr)\\
&& \times\Bigl[Y_{\ell,+1}(\theta' , \phi')+Y_{\ell,-1 }(\theta', \phi')\Bigr]\nonumber,
\end{eqnarray}
where $H_0=\sqrt{\epsilon_h\epsilon_0/\mu_0}\,E_0$ 
and $ j_{\ell}$ and $Y_{\ell,m }$ denote the
 spherical Bessel functions and the spherical harmonics, respectively~\cite{DLMF25.12}.

 To describe an incident plane wave propagating along a generic direction
 defined by the spherical angles $\theta_j$ and $\phi_j$, we 
 implement a rotation from the primed to the un-primed coordinate system with the help of 
 the Wigner rotation matrix elements $d^{\ell}_{m,\pm 1}(\theta_j)$ \cite{Edmonds1957}. 
 We then derive the Debye potentials for the incident field by taking a superposition of the plane-wave potentials (\ref{PiEPlane}) and  (\ref{PiMPlane})
 as in Eq.~(\ref{incidentfield}). The explicit expressions are written as sums over $\ell$ (for the total angular momentum $J^2$) and $m$ (corresponding to $J_z$) of the form
 \[
\sum_{\ell,m}\equiv  \sum^{\infty}_{\ell=1}\, \sum^{\ell}_{m=-\ell}.
 \]
 We find
 \begin{eqnarray}
\Pi_{\rm in}^{E}(r,\theta,\phi)&=& i\,\frac{E_{0}}{k} \sum_{\ell,m}\,   \gamma^{E}_{\ell,m}\, j_{\ell}(kr)\,Y_{\ell,m}(\theta,\phi)\\
\Pi_{\rm in}^{M}(r,\theta,\phi)&=& \frac{H_{0}}{k}\sum_{\ell,m}\,   \gamma^{M}_{\ell,m} \, j_{\ell}(kr)\, Y_{\ell ,m}(\theta,\phi),
\end{eqnarray}
The multipole coefficients $ \gamma^{E,M}_{\ell,m}$ are written as a sum over the plane waves $j=1,2$ and the photon helicity $\varepsilon=\pm 1:$
\begin{equation}
\gamma^{E,M}_{\ell m}=\sqrt{\frac{\pi(2\ell+1)}{\ell(\ell+1)}}i^\ell\, \sum_{j,\varepsilon} 
 \,g^{E,M}(\varepsilon)\,
e^{-i \phi_{j}(m-\varepsilon)}d^{\ell}_{m,\varepsilon}(\theta_{j})
\end{equation}
with $g^{E}(\varepsilon)=\varepsilon$ and $g^{M}(\varepsilon)=1.$

As we consider spherically-symmetric particles, it is straightforward to obtain the Debye potentials for the scattered field 
${\bf E}_{\rm s}$
by following the 
prescription outlined above: 
\begin{equation}\label{PiEs}
\Pi_{\rm s}^{E}(r,\theta,\phi)= i\frac{E_{0}}{k} \sum_{\ell,m}\,  (-a_\ell) \gamma^{E}_{\ell ,m} h^{(1)}_{\ell}(kr)Y_{\ell,m}(\theta,\phi)
\end{equation}
{\begin{equation}\label{PiMs}
\Pi_{\rm s}^{M}(r,\theta,\phi)= \frac{H_{0}}{k}\sum_{\ell,m}\,(-  b_\ell) \gamma^{M}_{\ell,m} h^{(1)}_{\ell}(kr)Y_{\ell,m}(\theta,\phi),
\end{equation}
where $h^{(1)}_\ell(k r) $ are the  spherical Hankel functions of the first kind~\cite{DLMF25.12}. 
The Mie coefficients
  $a_{\ell}$ and $b_{\ell}$ denote the scattering amplitudes
  for electric and magnetic multipoles, respectively.
 For a homogeneous dielectric sphere  embedded in a non-absorbing and non-magnetic host medium, they are given by~\cite{Bohren1983}
\begin{equation}\label{aell}
 a_{\ell}= \dfrac{n_s\psi_{\ell}(n_sx) \psi'_{\ell}(x)-\mu_s\psi_{\ell}(x) \psi'_{\ell}(n_sx)}{n_s\psi_{\ell}(n_sx) \xi'_{\ell}(x)-\mu_s\xi_{\ell}(x) \psi'_{\ell}(n_sx)} 
 \end{equation}
\begin{equation}
\label{bell}
 b_{\ell}=  \dfrac{\mu_s \psi_{\ell}(n_sx) \psi'_{\ell}(x)-n_s  \psi_{\ell}(x) \psi'_{\ell}(n_sx) }{\mu_s \psi_{\ell}(n_sx) \xi'_{\ell}(x)-n_s\xi_{\ell}(x)\psi'_{\ell}(n_sx), } 
\end{equation} 
where  $x=k R$ is the size parameter,
$n_s$ is the  relative refractive index of the sphere with respect to the host medium, 
 $ \mu_s$ is the relative magnetic permeability of the sphere, 
and
 $ \psi_{\ell}  $, $ \xi_{\ell} $ are the Riccati-Bessel functions \cite{DLMF25.12}.

Once the total field ${\bf E}={\bf E}_{\rm in}+{\bf E}_{\rm s}$ is known in terms of the Debye potentials for the incident and scattered fields, 
we are able to obtain the optical force acting upon the dielectric sphere by integration of the Maxwell stress tensor over a spherical surface at infinity:
\begin{equation}
{\bf F}= \lim_{r\rightarrow \infty}\left[ -\frac{r}{2} \int {\bf r}\left( \epsilon_h \epsilon_0 E^2 + \mu_0 H^2 \right) d\Omega  \right]. 
\label{Max}
\end{equation}

We choose our coordinate system with the $z$-axis bisecting the propagation directions as shown in fig.~\ref{fig:a1}: 
$\theta_1=\theta_2=\Theta,$ where $2\Theta$ is the angle between the two propagation directions. 
As the two plane waves have equal amplitudes and the same polarization,  the optical force points along the $z$ direction by symmetry.
$F_z$ has two distinct contributions:
the extinction term  $F_{\rm e}$ arises from cross terms of the form ${\bf E}_{\rm in}\cdot \textbf{E}^*_{\rm s}$ (and likewise for the magnetic field) and represents 
the rate of linear momentum removal from the incident fields.
Not all of this momentum is transferred to the particle, as part of it is carried away by the scattered fields. Thus, the second contribution to the optical force $F_{\rm s},$
which is  quadratic in  $\textbf{E}_s$ and   $\textbf{H}_s,$
represents the negative of the rate of momentum contained in the scattered electromagnetic fields.
 We find 
\begin{equation}
 F_z= F_{\rm s}+F_{\rm e}  \label{fz}
 \end{equation}
The scattering contribution is obtained by writing ${\bf E}_s$ and ${\bf H}_s$ in terms of the Debye potentials
 as in Eqs.~(\ref{EfromPi}) and (\ref{HfromPi}), respectively, and then taking the asymptotic 
approximation of the Hankel functions in (\ref{PiEs}) and (\ref{PiMs}) when evaluating the Maxwell stress tensor surface integral (\ref{Max}):
\begin{eqnarray} \label{Fs}
 F_{\rm s}  = \frac{\epsilon_h \epsilon_0 E_{0}^2}{k^2}\sum_{\ell,m}{\rm Im}\,\biggl\lbrace \ell(\ell+2) \sqrt{\frac{(\ell+1-m)(\ell+1+m)}{(2\ell+1)(2\ell+3)}}\nonumber\\
\times
\biggl[
 a_\ell a_{\ell+1}^*\gamma^{E}_{\scriptscriptstyle{\ell,m}}\gamma^{E^*}_{\scriptscriptstyle{\ell+1,m}} 
+ 
b_\ell b_{\ell+1}^*\gamma^{M}_{\scriptscriptstyle{l,m}}\gamma^{M^*}_{\scriptscriptstyle{\ell+1,m}}\biggr]
-im \,b_\ell\, a_\ell^*\gamma^{M}_{\scriptscriptstyle{\ell,m}}\gamma^{E^*}_{\scriptscriptstyle{\ell,m}}\biggr\rbrace 
 \end{eqnarray}
The extinction term results from interference between the incident and scattered field and reads
\begin{eqnarray}\nonumber
F_{\rm e}=\frac{\sqrt{\pi} \epsilon_h \epsilon_0 E_{0}^2}{2k^2}\,\sum_{\ell,m}\sum_{\varepsilon=\pm 1}
\varepsilon\,\sqrt{\ell(\ell+1)(2\ell+1)}\\
\times\,{\rm Re} \biggl[\biggl(a_\ell^*\gamma^{E^*}_{\scriptscriptstyle{\ell,m}}+\varepsilon
 b_\ell^*\gamma^{M^*}_{\scriptscriptstyle{\ell,m}} \biggr)
G^{\varepsilon}_{\ell,m} \biggr] \label{Fe}
 \end{eqnarray}
Eq.~(\ref{Fe}) contains an
additional sum over the photon helicity $\varepsilon=\pm 1$ and the associated  
 coefficients
\[
G_{\ell,m}^{\varepsilon}= i^\ell \sum_{j=1}^2\cos\theta_{j}\, d^{\ell}_{m,\varepsilon}(\Theta)e^{-i (m- \varepsilon)\phi_{j}}\,.
\]

\textit{Dipolar limit.}
In order to gain a more physical insight, we compare the limiting values of our 
exact analytical expressions when $kR \ll 1$ with the known results for the force on an induced dipole. 
Within the dipolar approximation, the electromagnetic response is entirely captured by the induction of electric and magnetic dipoles: 
${\bf p}=\epsilon_0 \alpha_e {\bf E}$ and ${\bf m}=\alpha_m {\bf H},$
 where $\alpha_e$ and $\alpha_m$ denote the electric and magnetic polarizabilities, respectively.
 
 Such model is obtained as a limiting case of Mie scattering for very small spheres, $kR\ll 1,$
 as long as the sphere is magnetic or provided that the sphere refractive index is high enough so as to satisfy the additional condition
 $n_s k R\gg 1$ \cite{Bohren1983}. In such cases,  the leading contribution in the extinction term (\ref{Fe}) arises from both electric and magnetic dipole terms $\ell=1.$ 
 The corresponding Mie coefficients are related to the electric and magnetic polarizabilities as follows:
\begin{equation}\label{abap}
a_1= -i\frac{k^{3}}{6 \pi}\, \alpha_e\;\;\;\;\;
b_1= -i\frac{k^{3}}{6 \pi}\, \alpha_m.
\end{equation} 

 Clearly, the extinction term (\ref{Fe}) cannot provide a pulling contribution as the total momentum removed from the incident 
 waves points along the positive $z-$ axis in fig.~\ref{fig:a1}. Indeed, the pulling force necessarily arises 
 from the scattered field carrying an excess linear momentum along the positive $z$-axis. 
Such effect is captured by the scattering contribution (\ref{Fs}), whose leading order term is proportional to $b_1a_1^*,$ representing the 
coherent interference between electric and magnetic dipoles \cite{Liu2017}. Together with the leading-order extinction terms, they 
lead to the dipolar force
 \begin{eqnarray}
 F_z & \approx & 2k \epsilon_h \epsilon_0 E^{2}_0  \Biggl[ {\rm Im} \Bigl( \frac{\alpha_e \cos\Theta}{\epsilon_h} + \alpha_m \cos^3\Theta\Bigr)\nonumber \\
&&
  - \frac{2 k^3}{3\epsilon_h } {\rm Re}(\alpha_e \alpha_m^*) \cos\Theta\Biggr] \;\;\;\;\mbox{(TE)} \label{dipTE}
\end{eqnarray}
in the case of TE-polarized waves, and to 
\begin{eqnarray} F_z & \approx & 2k\epsilon_h\epsilon_0 E^{2}_0   
  \Biggl[ {\rm Im} \Bigl( \frac{\alpha_e \cos^3\Theta}{\epsilon_h}+  \alpha_m \cos\Theta\Bigr)\nonumber \\
&&
  - \frac{2 k^3}{3\epsilon_h } Re(\alpha_e \alpha_m^*) \cos\Theta\Biggr]  \;\;\;\;\mbox{(TM)} \label{dipTM}
\end{eqnarray} 
in the case of TM polarization. Such expressions are obtained by neglecting the quadrupole and higher multipoles ($\ell\ge 2$) in  (\ref{Fs}) and (\ref{Fe})
 and using the explicit form of the Wigner matrix elements $d^{1}_{m,\pm 1}(\Theta).$
They agree with known results \cite{Liu2017,Mobini2018} for the dipolar regime. 

In the case of very small non-magnetic microspheres with moderate refractive indexes, the magnetic dipole turns out 
to be much smaller than the electric dipole, and actually comparable to the electric quadrupole term associated to $a_2$ (Rayleigh scattering regime)~\cite{Bohren1983}. 
Thus,  OPF cannot be achieved in the Rayleigh limit, as the electric-magnetic dipole interference term appearing in (\ref{dipTE}) and (\ref{dipTM})
would be missing. 

 In the next section, we discuss in more detail the validity of the dipolar approximation, by comparing the full evaluation of the Mie series expressions
 (\ref{Fs}) and (\ref{Fe}) with the approximations (\ref{dipTE}) and (\ref{dipTM}) consisting in keeping only the dipolar terms involving $a_1$ and $b_1.$ 
 We will confirm that the dipolar approximation is more accurate for higher refractive indexes. As expected, this is also the case allowing
 to achieve OPF with smaller particles. 

\section{Results and discussions } 

For the numerical examples discussed in this section, we take
 the vacuum wavelength to be $\lambda_0=1064\,{\rm nm}.$ 
As the host medium, we consider an aqueous solution with $n_h=1.332.$
We normalize the optical force to $F_{0}= 2 \pi  n_h  I_0/(k^2 c)$, where $I_0=\sqrt{\epsilon_h\epsilon_0/\mu_0}\, E_0^2/2$ is the intensity of 
each incident plane wave  shown in Fig.~\ref{fig:a1}.
 
 In fig.~\ref{fig:2}, we  compare the exact results  for the 
 optical force (solid line), calculated from Eqs. (\ref{fz}), (\ref{Fs}) and (\ref{Fe}), with the  dipolar  approximations
  (\ref{dipTE}) and (\ref{dipTM}) (dashed line). For the latter, the polarizabilities are calculated  from
 (\ref{abap}) by taking the full exact expressions for the Mie coefficients $a_1$ and $b_1.$

In figures \ref{fig:2}(a)-(b), we consider a silicon sphere
(refractive index $3.5$) of radius $R=140\,{\rm nm}$ and plot the axial force 
 as a function of the half angle $\Theta$ between the two incident wavevectors as shown in Fig \ref{fig:a1} for (a) TE and (b) TM polarizations. 
 The figures show that the dipolar approximation is capable of reproducing the features of the exact curve in this example with a small sphere and high refractive index. 
OPF is achieved in the interval  $65\degree<\Theta<90\degree$ only in the case of TE polarization and is explained by the coherent interference between 
electric and magnetic dipoles as discussed in connection with Eq.~(\ref{dipTE}).
The maximum pulling force 
   occurs at $\Theta = 78^{\rm o}.$ 
    In Figs.~\ref{fig:2}(c)-(e)
    we fix  the incident angle at $ \Theta= 78^{\rm o}$
     and plot the force as function of the sphere radius $R.$
  Figs. \ref{fig:2}(c)-(d)  correspond to Si microspheres, whereas 
  \ref{fig:2}(e)-(f) show the results for SiO$_2$ microspheres, with refractive index $ n_{\rm SiO_2}=1.45.$
 The polarization is TE in (c) and (e) and TM in  (d) and (f).

 \begin{figure} 
\includegraphics[width = 3. in]{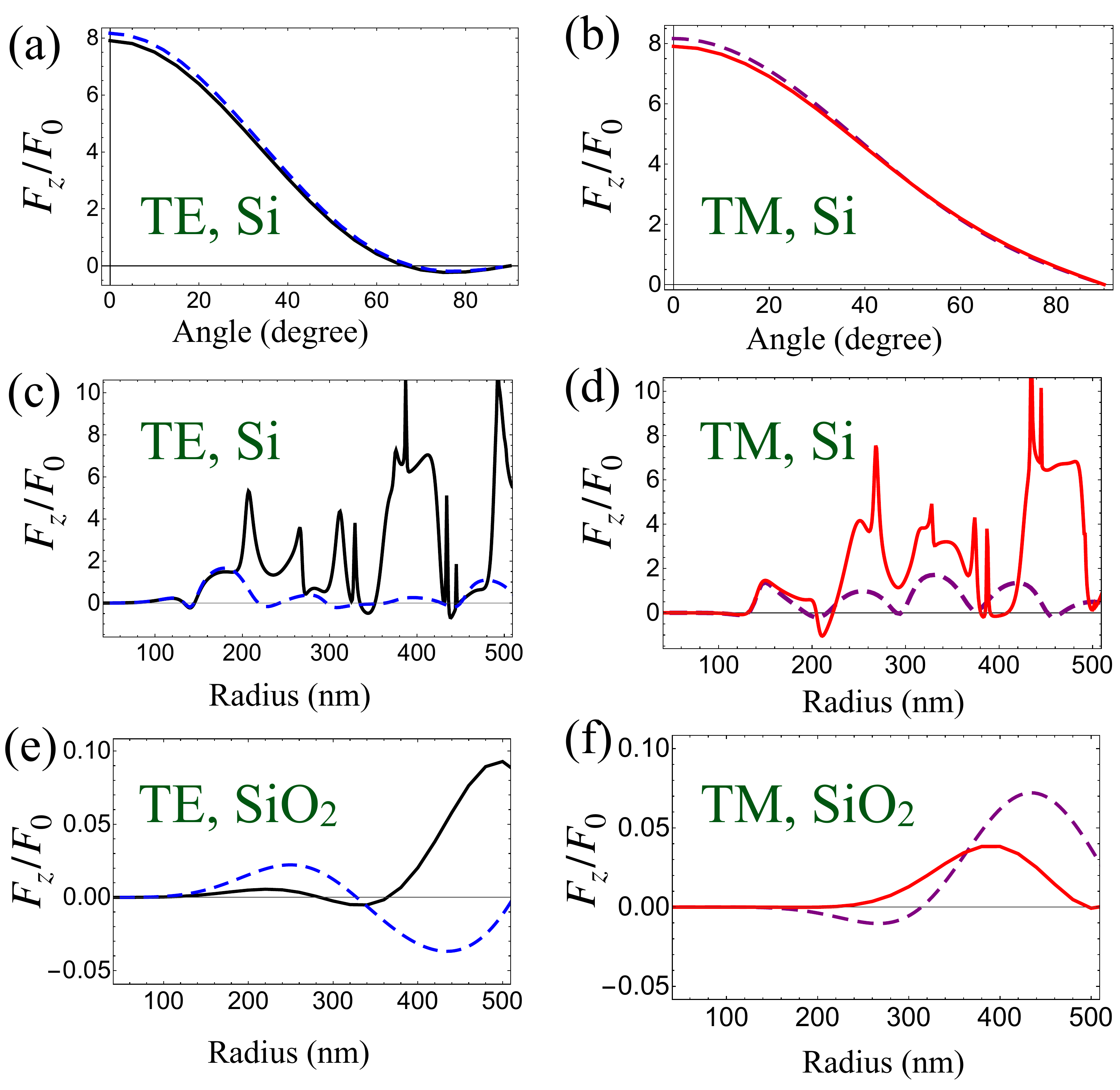}
\caption{ Normalized optical force acting on Si 
or   SiO$_2$  microspheres  under illumination by two plane waves with the same intensity and linear polarization. 
All panels show the result of the full Mie calculation (solid) and of the dipolar approximation (dashed). 
The polarization is transverse electric (TE) in (a,c,e) and transverse magnetic (TM) in (b,d,f)  (see fig.~\ref{fig:a1}). 
(a,b) Optical force as a function of the half angle $\Theta$ for a Si
microspheres of radius  $140\,{\rm nm}.$ 
Panels (c,d) for Si and (e,f) for SiO$_2$ show the optical force as a function of sphere radius for
$\Theta = 78^{\rm o}.$}  
\label{fig:2}
\end{figure}
 
 In the case of silicon, OPF is also achieved for TM polarization as shown in Fig.~\ref{fig:2}(d). In contrast with the TE configuration, the pulling effect here 
 is not captured by the dipolar curve and 
 is then not related to interference between electric and magnetic dipoles. Instead, it results from  contributions of higher multipoles
 yielding an enhancement of scattering along the 
 $z$-axis bisecting the directions of incidence
and  which can only
 be understood within the full Mie theory as it takes place at larger radii. 

 As expected, the dipolar approximation fails to describe the behavior of the optical force on 
 SiO$_2$ microspheres in the size range shown in Figs. \ref{fig:2}(e)-(f). Since its refractive index is lower than in the case of silicon, 
very small SiO$_2$ spheres
 behave as induced electric dipoles with a negligible magnetic dipole contribution. As a consequence, the exact optical force is mostly positive in the range
shown in \ref{fig:2}(e)-(f) (except for a negligible pulling effect near $R\sim 320\,{\rm nm}$ for TE polarization). 
 In short, OPF is not found in the case of very small SiO$_2$ beads as the scattering angular distribution resembles the electric dipole (Rayleigh) 
 distribution and hence does not favor the forward direction.  

To circumvent this issue we put forward the strategy of doping the SiO$_2$ spheres 
with gold spherical inclusions (radius $a$) in order to excite electric and magnetic dipoles simultaneously.
We assume $a\ll \lambda$ and then model 
 this system 
 as an effectively homogenous medium with the help of the extended Maxwell-Garnet theory \cite{Ruppin2000,Mallet2005}. The effective refractive index
 of the composite sphere 
$n_{\rm eff}=\sqrt{\epsilon_{\rm emg} \mu_{\rm emg}}$ is obtained from the effective relative permittivity and permeability:
\begin{equation}
\epsilon_{\rm emg}=\epsilon_{\rm h}\dfrac{x_i^3+3if a^i_1}{x_i^3-\frac{3}{2}if a^i_1} 
\label{emg}
\end{equation}
\begin{equation}
\mu_{\rm emg}=\dfrac{x_i^3+3if b^i_1}{x_i^3-\frac{3}{2}if b^i_1}
\label{muemg}
\end{equation}
where $f$ denotes the volume filling fraction. The dipolar Mie coefficients 
of the  inclusions $a^i_1$ and $b^i_1$ are derived from the corresponding 
size parameter
 $x_i = n_{\rm SiO_2}\omega a/c$ and from the value 
  $\epsilon_{\rm Au}=-48.45+3.6\,i$
 for the gold relative permittivity at $\lambda_0=1064\,{\rm nm}$ \cite{Johnson1972}. 

In Fig.~\ref{fig:4}, we plot 
the optical  force  as a function of sphere radius $R$
for TE (black) and TM (red) polarizations. We take $\Theta = 85^{\rm o}$ as the half-angle between the two beams. 
The results for  homogenous SiO$_2$ spheres ($f=0$) are shown in \ref{fig:4}(a), whereas
the force on composite spheres with gold inclusions ($f=0.18$) are shown in \ref{fig:4}(b).
Measurable OPF on homogeneous SiO$_2$ spheres can be obtained for relatively large radii
in the case of TE polarization, while the force is always positive for TM. Such striking difference between the two cases shows that the relative 
phase between the scattered fields produced by each plane wave component strongly depends on the incident polarization. Thus, the condition for constructive interference 
near the forward direction can be controlled by the incident polarization. 

 \begin{figure} 
 \centering
\includegraphics[width = 3.4 in]{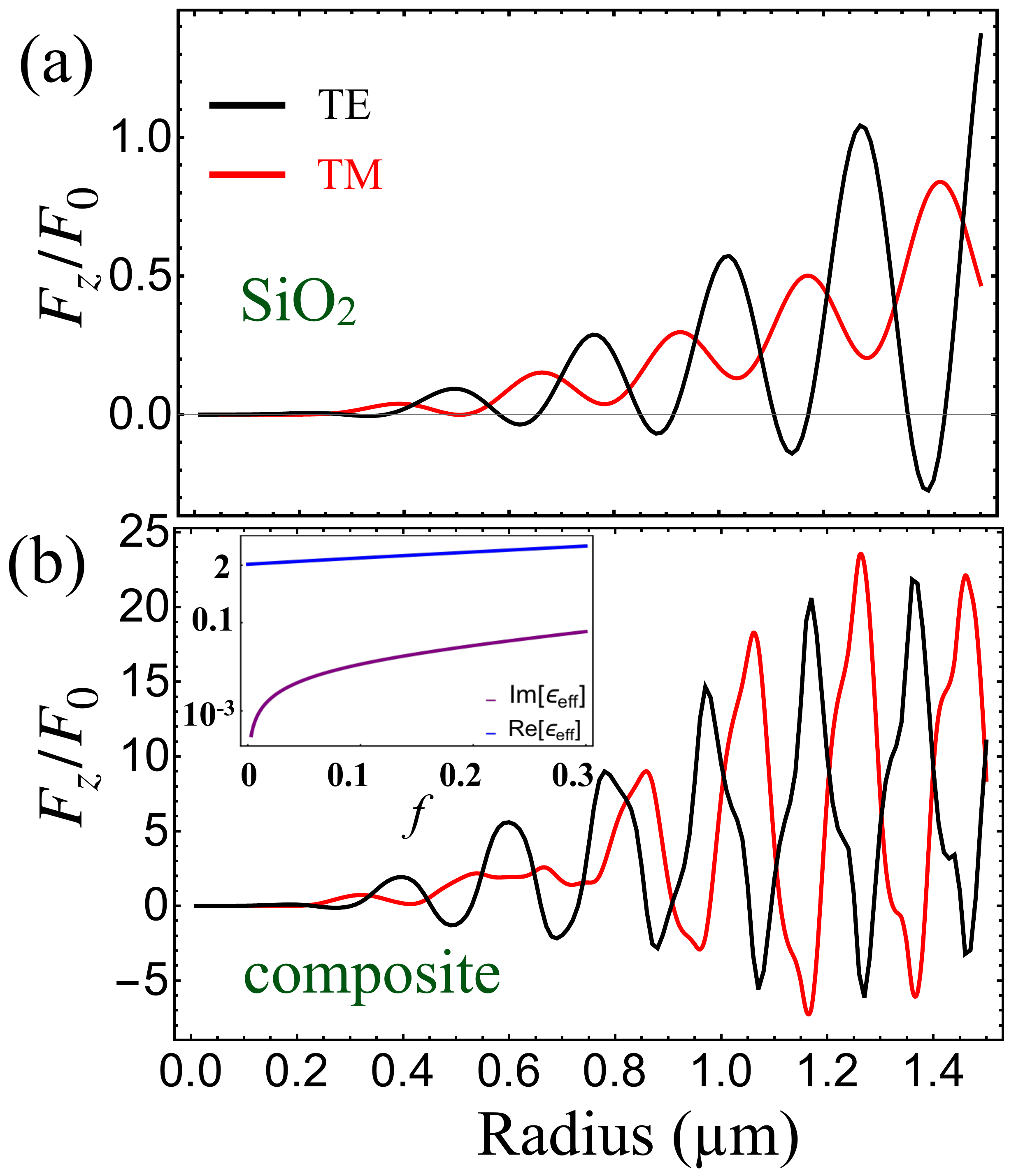}
\caption{ Normalized optical force  as a function of  sphere radius  for   (a)  homogeneous SiO$_2$ microspheres and  (b) composite SiO$_2$ microspheres with gold inclusions 
(filling fraction $f=0.18$).  The incident plane waves are either 
 TE= (black) or TM-polarized (red). The half angle between the incidence directions is $\Theta=85^{o}.$ 
 The inset shows the real (blue) and imaginary (violet) parts of the composite effective permittivity $\epsilon_{\rm eff} $ as a function of the filling fraction. }
\label{fig:4}
\end{figure}

The scenario is drastically changed when one considers the  SiO$_2$ host sphere with gold inclusions, as shown in Fig. \ref{fig:4}(b). 
 Indeed, the presence of gold inclusions not only allows for OPF for
 both TE and TM polarizations, but also increases its magnitude by about one order of magnitude. 
 At specific size ranges and depending on the polarization, the fields scattered by the inclusions interfere constructively (destructively) near the forward (backward) direction, 
 then leading to a strong pulling effect. The resulting optical forces for TE and TM polarizations oscillate
 nearly out-of-phase as a function of the radius as shown in  Fig. \ref{fig:4}(b). 
 Thus, the size intervals allowing for OPF using TE beams is approximately the complement of 
 the intervals allowing for OPF using TM beams. Such feature indicates the possibility of a polarization-controlled particle sorting according to particle size. 
 
 Adding metallic inclusions also allows to extend the range of optical pulling towards smaller sizes.
Indeed, the effective permittivity $\epsilon_{\rm emg}$ increases with the filling fraction $f,$ as shown in the inset of fig.~\ref{fig:4}(b).
Moreover, the inclusions also lead to an effective permeability $\mu_{\rm emg}$ slightly different from one
according to Eq.~(\ref{muemg}). 
 In line with the discussion of sec.~2, both effects enhance the magnetic dipole contribution, which is essential for 
 achieving OPF on small particles. 
 
The occurrence of OPF on small particles is better visualized in the density plot of the OPF versus sphere radius and filling fraction shown in 
Fig.~\ref{fig:5}, for (a) TE and (b) TM polarizations. 
Colored areas indicate optical pulling forces whereas the regions corresponding to optical pushing forces are left blank for clarity. 
The overall disposition of the colored regions indicates that  
the conditions for OPF become more selective for bigger particles as a larger number of multipoles contribute to scattering. Indeed, 
in this case a fine tuning of the material parameters is required to achieve the simultaneous interferometric conditions involving all multipoles contributing to the 
scattering force component (\ref{Fs}). 

In the case of TE polarization, the colored areas appear as a pattern of  stripes 
illustrating how the 
 size intervals allowing for OPF depend on the filling fraction.
Within a given stripe, the magnitude of the OPF is initially enhanced as $f$ increases from zero (homogenous case). 
Stripes corresponding to larger radii (upper part) are increasingly more inclined and tend to shrink as the filling fraction increases. Both features are related to 
light absorption by the microsphere. 
The inset of fig.~\ref{fig:4}(b) shows that ${\rm Im}(\epsilon_{\rm eff})$ increases sharply as a function of $f.$
As the penetration depth $\delta=\lambda_0/[4\pi \,{\rm Im}(n_{\rm eff})]$ decreases approaching the sphere diameter $2R,$ the fraction of absorbed light increases, which is clearly detrimental to the pulling effect. For very small spheres, $R\ll\delta,$ absorption is still negligible
and then the width of the lower stripes in Fig.~\ref{fig:5}(a) are approximately uniform. 
On the other hand, the stripes corresponding to larger radii are more affected by absorption as expected. They shrink and eventually
 disappear as $f$ and consequently ${\rm Im}(n_{\rm eff})$ increase.
Also, the inclination of the stripes shows that the reduction of the penetration depth $\delta$ is compensated by a sharp decrease of the radii allowing for pulling so as to keep
the ratio $R/\delta$ approximately unchanged. 

The overall picture is similar in the case of TM polarization shown in Fig.~\ref{fig:5}(b). However, the pulling regions are more scarce, and very small 
filling fractions are excluded in this case.

\begin{figure} 
\includegraphics[width = 3.4 in]{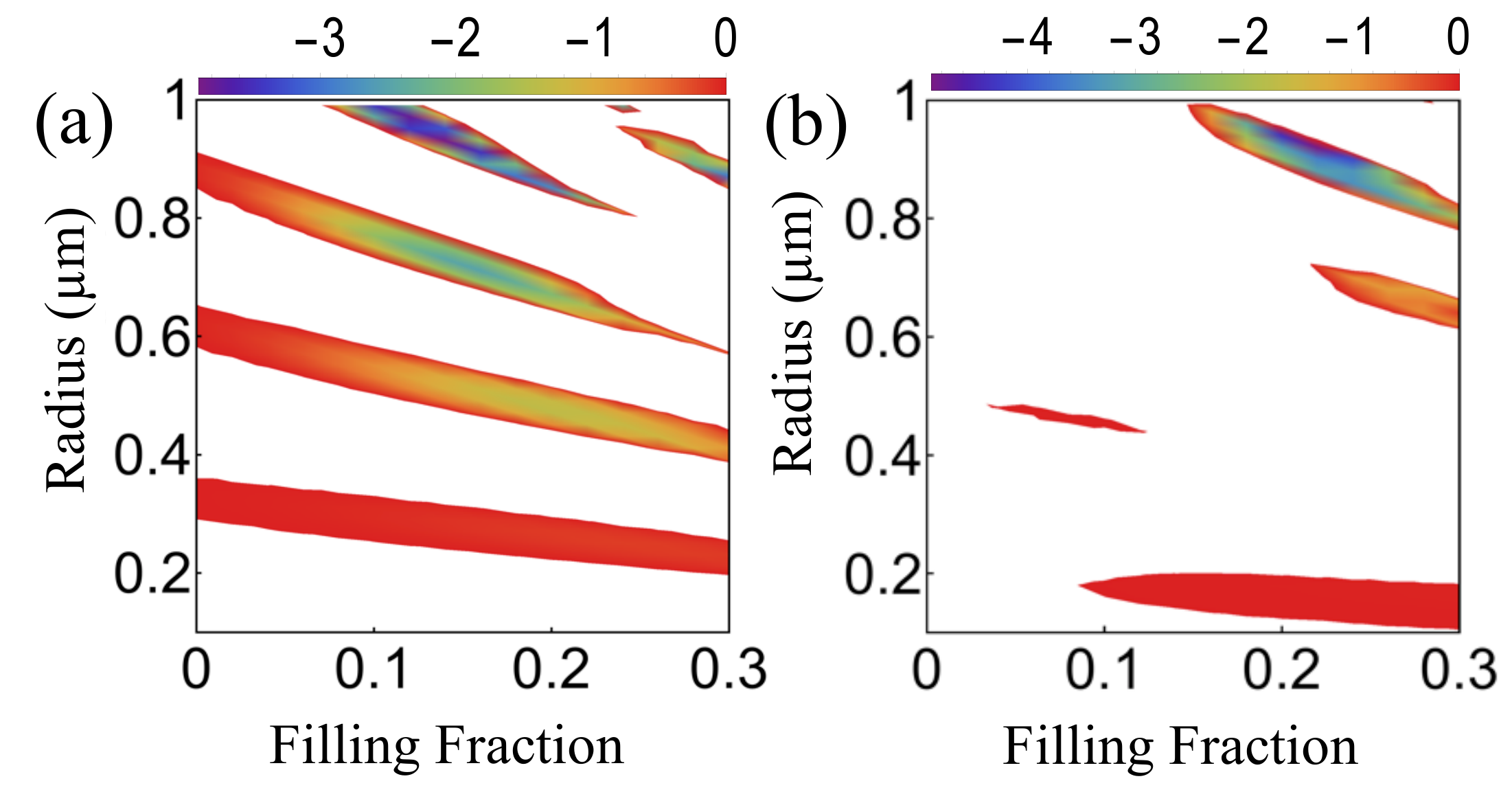}
\caption{ Optical pulling force on a composite microsphere 
as function of radius and  filling fraction. For clarity, only regions in the parameter space leading to pulling forces are shown. 
 (a) transverse electric (TE) and (b) transverse magnetic (TM) polarization. The  half-angle between the incidence directions is fixed at $\Theta=85^{\rm o}.$ }
\label{fig:5}
\end{figure}

\section{Conclusion}

We have developed  a theoretical framework to calculate the optical pulling force 
on a microsphere illuminated by a superposition of plane waves. Due to the rotational symmetry of the scatterer, Mie scattering of plane waves propagating along arbitrary
directions can be connected with the more standard case of axial propagation by employing Wigner rotation matrices and Debye potentials. We have derived an explicit 
result for the optical force as a partial-wave series when considering the example involving two linearly polarized 
plane waves. However, it is straightforward to extend our approach to multiple plane waves with arbitrary polarizations. 
The case of circular polarization might be particularly interesting given its possible application on enantioselective manipulation of chiral particles.

Our results show that TE-polarized waves can pull  low and high refractive-index particles alike. In the former case, the technique is limited to larger sizes, 
involving higher multipoles beyond the electric and magnetic dipoles. On the other hand, TM-polarized waves allow  for OPF only on high-refractive index particles.
We have shown that the use of a metamaterial platform  not only
 leads to an order of magnitude increase in the pulling force, but also allows to
pull smaller particles, thus extending the technique into the dipolar regime. 
As a specific example, we have considered a low-index dielectric sphere doped with plasmonic inclusions. 
The strong enhancement in the pulling force is achieved for small values of the filling fraction,  to avoid the detrimental effect of absorption. 
In this configuration, the size ranges which allow for pulling for each orthogonal polarization are approximately complementary. Such feature could be 
applied to implement a polarization-controled particle sorting using OPF.

\textbf{Funding.} National Council for Scientific and Technological Development (CNPq--Brazil),
Coordination for the Improvement of Higher Education
Personnel (CAPES--Brazil), the National Institute of Science
and Technology Complex Fluids (INCT-FCx), and the
Research Foundations of the States of Minas Gerais
(FAPEMIG), Rio de Janeiro (FAPERJ) and S\~ao Paulo
(FAPESP).

\textbf{Acknowledgment.} We thank Felipe S. S. da Rosa for inspiring discussions. 
%\section*{References}

% Bibliography
%\bibliography{sample}

\end{document}